\documentclass[10pt,twocolumn,aps,prx,floatfix,superscriptaddress]{revtex4-2}
\usepackage{graphicx}
\usepackage{color}
\pdfoutput=1
\usepackage{amsmath}
\usepackage{amssymb}
\usepackage{comment}
\usepackage{hyperref}

\begin{document}
\title{Pressure-Induced B1 to B2 Phase Transition in CeN Studied by \textit{ab initio} Correlation Matrix Renormalization Theory Calculations}
\date{\today}

 \author{Jun Liu}
\affiliation{Ames Laboratory --USDOE, Iowa State University, Ames, IA 50011}

 \author{Jianhua Zhang}
 \email{jianhuazhang@hainanu.edu.cn} 
 \affiliation{School of Physics and Optoelectronic Engineering, Hainan University, Haikou, China}
\affiliation{Department of Physics and Astromany, Iowa State University, Ames, IA 50011}

 \author{Yongxin Yao}
 \affiliation{Ames Laboratory --USDOE, Iowa State University, Ames, IA 50011}
 \affiliation{Department of Physics and Astromany, Iowa State University, Ames, IA 50011}
 \author{Kai-Ming Ho}
 \affiliation{Ames Laboratory --USDOE, Iowa State University, Ames, IA 50011}
 \affiliation{Department of Physics and Astromany, Iowa State University, Ames, IA 50011}
 
 \author{Cai-Zhuang Wang}
  \email{wangcz@ameslab.gov} 
\affiliation{Ames Laboratory --USDOE, Iowa State University, Ames, IA 50011}
\affiliation{Department of Physics and Astromany, Iowa State University, Ames, IA 50011}

\begin{abstract}
We apply correlation matrix renormalization theory (CMRT) to cerium nitride (CeN) under pressure. For B1 (NaCl-type) phase, CMRT gives an equation of state consistent with ambient pressure experiments. It produces electronic density-of-state (DOS) characterized by a sharp $4f$ quasi-particle resonance peak pinned at the Fermi level
and two subbands formed by strong hybridization between the localized Ce-$4f$ electrons and the itinerant Ce-$5d$ and N-$2p$ electrons below the Fermi level, consistent with XPS experiments. Upon compression, CMRT predicts a first-order B1 → B2 (CsCl-type) transition with ${\sim}$11\% volume collapse in agreement with experiments. Across the transition, the $4f$ spectral weight broadens, the $4f$ orbital occupancy increases, and the hybridization with conduction states enhances, signaling a crossover from
partially localized to more itinerant $4f$ behavior. These features are in excellent agreement with experimental observations, demonstrating that CMRT provides a parameter-free description and prediction of correlation-driven structural and electronic transitions in rare-earth compounds.
\end{abstract}

\maketitle

\section{Introduction}

Cerium-based compounds have drawn significant scientific attention due to their rich and exotic physical properties, which originate from the complex behavior of Ce $4f$ electrons \cite{Chen2018,Lou2021,Wu2021,Aynajian2012}. These electrons exhibit a dual nature---being both localized and itinerant---leading to strong electron correlation effects and intricate interactions with conduction electrons. As a result, cerium-based compounds often exhibit novel quantum phenomena, such as the Kondo effect \cite{Lou2021,Christovam2024}, heavy-fermion behavior \cite{Aynajian2012}, superconductivity \cite{Squire2023,Chajewski2024,Chen2024}, mixed-valence states \cite{Baer1977,Baer1978,Schlegel1978,Olcese1979}, and quantum phase transitions \cite{Chajewski2024,Dzsaber2017}. Among cerium-based compounds, cerium nitride (CeN) stands out due to its unusually small lattice constant. While most cerium-based compounds show localized $4f^{1}$ states resembling $\gamma
$-Ce, CeN exhibits partial itinerant $4f$ behavior similar to $\alpha $-Ce, leading to a mixed-valence ground
state \cite{Litsarev2012,Svane1998}. High-energy photoemission (XPS and BIS) consistently reveals intermediate valence, highlighting strong
valence fluctuations and partial delocalization of $4f$ electrons \cite{Wuilloud1985}.

CeN is characterized by its anomalous mixed-valence ground state, which arises from the dual localized–itinerant nature
of Ce $4f$ electrons \cite{Lu2020,Zhao2025}. Multiple experiments and theoretical calculations have confirmed this intermediate-valence
behavior. XPS measurements confirm this anomalous behavior, showing a mixed valence of $\sim3.46$ at room temperature and
suggesting that the B1-phase of CeN corresponds to an intermediate-valence state \cite{Baer1977}. Optical reflectivity and Drude analysis further reveal an intermediate valence of about 3.4 at 300 K \cite{Schlegel1978}. Lattice-parameter-based estimates gave
$\sim3.85$ at room temperature, decreasing toward purely trivalent above 1100 K \cite{Olcese1979}. High-energy spectroscopy
experiments (XPS and BIS) found that the effective valence is close to $3^{+}$, with the $4f^{1}$ state lying near the conduction
band edge and showing valence fluctuation signatures \cite{Wuilloud1985}. More recently, Diniz re-evaluated
Olcese$'$s estimate and demonstrated that the Ce$^{4+}$ fraction is much smaller (as low as
$\sim18\%$,  $n_f \approx 3.18$ ) once charge transfer and $4f$–conduction hybridization are included \cite{Diniz2020}. Recent DFT+DMFT (density functional theory with dynamical mean-field theory)
calculations reveal that B1-type CeN is a mixed-valence ( $n_f \approx 0.89$) system at ambient pressure, but under
compression or tension it undergoes successive quantum phase transitions from a mixed-valence state to a Kondo metal
and eventually to an insulator \cite{Liu2024}. However, some researchers have argued that the two valence configurations 
(tetravalent $4f^{0}$ and trivalent $4f^{1}$) are not degenerate, making mixed valence in the ground state of CeN highly
unlikely \cite{Kanchana2011}.

The high-pressure behavior of cerium nitride  has attracted considerable attention due to the strong interplay between its crystal structure and $4f$-electron states. At ambient conditions, CeN crystallizes in the NaCl-type (B1, Fm$\bar3$m) structure. Upon compression, CeN exhibits a structural phase transition from the B1 to a high-pressure phase.
Energy-dispersive X-ray diffraction and synchrotron measurements using diamond anvil cells reported the onset of the B1→B2 (CsCl-type, Pm$\bar3$m) transition at 65–70 GPa with a relative volume collapse of about 11\% \cite{Olsen2012}. In contrast, other
cerium monopnictides CeX (X=P \cite{Vedel1987}, As \cite{Werner1983},Sb \cite{Leger1984},
	Bi\cite{Leger1985}) undergo the same transition at significantly lower
	pressures (11–19 GPa). This systematic decrease in transition pressure with increasing pnictogen size can be
attributed to their larger lattice parameters and correspondingly weaker Ce–X bonding strength. The more compact
lattice of CeN requires higher compression to stabilize the denser B2 phase. These observations highlight the close
correlation between lattice constants and transition pressures in Ce-based pnictides. Studying
CeN$'$s B1–B2 transition is important because it provides insight into how pressure-driven changes
in bonding and Ce--N hybridization correlate with $4f$ delocalization. In particular, the collapse into the high-pressure
B2 phase is thought to involve substantial delocalization of the Ce \leavevmode\,$4f$ electron, and thus offers a concrete
example of a correlation-driven structural transition in a heavy-fermion compound.

Given these experimental findings on CeN$'$s pressure-induced B1–B2 transition, a wide range of
theoretical approaches have been employed to clarify its electronic structure and phase stability. For example,
calculations using generalized gradient approximation (GGA) in the density functional theory (DFT) suggested itinerant
$4f$ electrons and predicted a B1–B2 transition near $\sim68$ GPa \cite{Olsen2012,Bhardwaj2016}, while self-interaction correction based on local
spin density approximation (SIC–LSDA) gave $\sim62$ GPa \cite{Svane1998}. A relatively simple improved interaction potential model yielded a higher estimate of $\sim88$ GPa \cite{Dubey2015}. GGA calculations also
proposed the possible existence of an intermediate B10 phase between 53 GPa and 200 GPa before reaching the B2 structure \cite{Sahoo2013}. More correlated extensions such as LDA+U and LSDA+U shifted the predicted transition pressure to lower values, around 40–52 GPa \cite{Zhang2013,Oualdine2018}, but still treated 4f electrons in a mean-field manner. Beyond these static approximations, dynamical mean-field theory has been employed to incorporate many-body correlations and valence fluctuations. LDA/GGA+DMFT studies found a somewhat higher transition pressure, around $\sim90$ GP, consistent with the view that the B1–B2 collapse is accompanied by enhanced delocalization of Ce $4f$ electrons \cite{Li2023}. Moreover, stress-dependent DMFT studies revealed that B1-type CeN undergoes successive quantum phase transitions from a mixed-valence state to a Kondo metal and eventually to an insulator \cite{Liu2024}, highlighting the essential role of dynamical correlations. Altogether, the reported transition pressures span a wide range (40–90 GPa), reflecting the extreme sensitivity of CeN$'$s structural stability to the treatment of $4f$ correlations and underscoring both the uniqueness of CeN within the CeX family and the limitations of static DFT-based descriptions.

Despite these efforts, each theoretical framework has intrinsic limitations. Standard DFT fails to capture 
the strong correlation and configuration mixing of $4f$ electrons; DFT+U improves localization but relies on an adjustable Hubbard U, leading to strong parameter dependence; and even DFT+DMFT, while accounting for valence fluctuations and many-body correlations, still depends on effective U and J values, double-counting schemes, and neglects intersite correlations beyond the single-site level. These issues complicate predictive consistency and partly explain the wide spread of reported transition pressures.

To overcome these challenges, we turn to the \textit{ab initio} correlation matrix renormalization theory (CMRT) recently developed \cite{Liu2021JPCM}. Unlike mean-field or parameterized approaches, CMRT constructs and solves the many-body equation of states directly, evaluating all Coulomb and exchange integrals among basis orbitals without empirical inputs \cite{Liu2022}. This parameter-free approach has already demonstrated its predictive power in $f$-electron systems. In cerium metal, CMRT clarified the essential role of Coulomb interaction in the phase formation of fcc Ce, correctly reproducing the energetics of the $\alpha $–$\gamma $ transition and identifying their distinct correlation characters via local $4f$ charge fluctuations \cite{Liu2021prb}. The method was subsequently applied to fcc Pr, where it achieved excellent agreement with experimental pressure–volume data and consistently described the correlation-driven evolution of $4f$ electrons in comparison with other state-of-the-art approaches such as LDA+DMFT \cite{Liu2022}. Building on these advances, the present work represents the first successful extension of CMRT to $4f$-electron compound systems. Specifically, we apply CMRT to CeN under pressure, aiming at avoiding the technical ambiguities of previous DFT and DFT+DMFT studies and to providing a parameter-free understanding of the interplay between lattice collapse and $4f$ delocalization with correlation effects included into the study.

\section{Methods}
The Correlation Matrix Renormalization Theory is a fully \textit{ab initio} variational approach designed for strongly correlated electron systems, employing a multiband Gutzwiller wave function as the trial ground state \cite{Liu2021JPCM}. In particular, CMRT provides a unified framework capable of treating both itinerant and localized electrons on the same footing within an electronic structure calculation, similar to DFT-based embedded correlated \textit{ab initio} methods \cite{Borghi2014} but avoiding using adjustable energy parameters to handle correlation effects.

For a periodic bulk system, the CMRT ground-state total energy is
	\begin{widetext}%
		\begin{align}
			E_{total}  &  =\sum_{\substack{ij\\\alpha\beta,\sigma}}\widetilde t_{i\alpha,j\beta
			;\sigma}\left\langle c_{i\alpha\sigma}^{\dag}c_{j\beta\sigma}\right\rangle +\dfrac
			{1}{2}\times\sum_{\substack{ijkl\\\alpha\beta\gamma\delta,\sigma\sigma^{\prime}%
			}} \widetilde U_{ijkl;\sigma\sigma^{\prime}}^{\alpha\beta\gamma\delta}  \left(  \left\langle c_{i\alpha\sigma}^{\dag}%
			c_{k\gamma\sigma}\right\rangle \left\langle c_{j\beta\sigma^{\prime}}^{\dag
			}c_{l\delta\sigma^{\prime}}\right\rangle -\delta_{\sigma\sigma^{\prime}%
			}\left\langle c_{i\alpha\sigma}^{\dag}c_{l\delta\sigma^{\prime}}\right\rangle
			\left\langle c_{j\beta\sigma^{\prime}}^{\dag}c_{k\gamma\sigma}\right\rangle
			\right) \nonumber\\
			& +E_{\mathit{local}} %
			  \label{ecmrB}%
		\end{align}
	\end{widetext}
with the dressed hopping and two-body interactions defined as 
\begin{align}
 \widetilde t_{\mathit{i\alpha },\mathit{j\beta };\sigma }=t_{\mathit{i\alpha },\mathit{j\beta }}+\frac{N_e} 2\lambda
_{\mathit{ijji};\sigma \sigma ^{'}}^{\mathit{\alpha \beta \beta \alpha }}, 
\end{align}
\begin{align}
\widetilde U_{\mathit{ijkl};\sigma \sigma ^{'}}^{\mathit{\alpha \beta \gamma \delta
}}=U_{\mathit{ijkl}}^{\mathit{\alpha \beta \gamma \delta }}-\lambda _{\mathit{ijkl};\sigma \sigma ^{'}}^{\mathit{\alpha
		\beta \gamma \delta }},
\end{align}
and the local energy  $E_{\mathit{local}}$ expressed as 
	\begin{align}
	E_{\mathit{local}}=\sum _i\sum _{\Gamma }\widetilde E_{i\Gamma }\left(p_{i\Gamma }-p_{i\Gamma _0}\right).
    \end{align}
Here,  $i,j,k,l$ denote site indices, while  $\alpha ,\beta ,\gamma ,\delta $ represent orbital indices, and  $\sigma
,\sigma '$ correspond to spin indices.  $\Gamma $ labels the Fock states in the occupation number representation of
local correlated orbitals on each atom within the unit cell, and  $N_{e}$ is the number of electrons per unit cell. The energy parameters  $t_{\mathit{i\alpha },\mathit{j\beta }}$ and  $U_{\mathit{ijkl}}^{\mathit{\alpha
		\beta \gamma \delta }}$ refer to the bare hopping and Coulomb integrals, respectively. To improve the accuracy of the
total energy calculation, CMRT introduces the sum rule correction term with its coefficient denoted as  $\lambda
_{\mathit{ijkl};\sigma \sigma ^{'}}^{\mathit{\alpha \beta \gamma \delta }}$. The quantities  $\widetilde E_{i\Gamma }$
\ represent the energies of the Fock states for the dressed local correlated Hamiltonian at each site.

The first two terms in Eq. (\ref{ecmrB}) represent the expectation value of the dressed lattice Hamiltonian within the
CMRT framework. In this context, the expectation values of two-body operators are expanded with the
Wick$'$s theorem and expressed in terms of one-particle density matrices, defined as follows:
\begin{align}
	\left\langle c_{i\alpha\sigma}^{\dag}c_{i\beta\sigma}\right\rangle  &
	=f\left(  z_{\alpha\sigma}\right)  f\left(  z_{\beta\sigma}\right)
	\left\langle c_{i\alpha\sigma}^{\dag}c_{i\beta\sigma}\right\rangle
	_{0}\nonumber\\
	&  +\left[  1-\delta_{\alpha\beta}f^{2}\left(  z_{\alpha\sigma}\right)
	\right]  \bar n_{i\alpha\sigma}.
\end{align}
Here,  $z_{\mathit{\alpha \sigma }}$ denotes the Gutzwiller renormalization factor, while $\left\langle \ldots\right\rangle _{0}$ represents the one-particle noninteracting density matrix, and 
$\bar n_{\mathit{i\alpha \sigma }}$\ \hspace{0pt}indicates the local electronic occupation of state  $\alpha $.
The function  $f(z_{\mathit{\alpha \sigma }})$ is introduced to anchor CMRT to the solution of an exactly solvable
model in a certain dimer limit \cite{Zhao2018}.

The third term in Eq. (\ref{ecmrB}) plays a crucial role in preserving the dominant local physics within CMRT. It rigorously expresses the local correlated energy via the variational parameter  $P_{i\Gamma}$\hspace{0pt}, which represents the occupation probability of Fock state  $\Gamma $ constructed from the correlated atomic orbitals at site $i$. Its noninteracting counterpart,  $P_{i\Gamma _0}$\hspace{0pt}, corresponds to the same quantity evaluated within a mean-field approximation and acts as a rigorous double counting term cancelling out the local energy components already considered in the first two terms of Eq. (\ref{ecmrB}). The local correlated Hamiltonian underlying the third term primarily includes two-body Coulomb interactions, which govern key local spin and charge physics. Its exact treatment enables the preservation of intrinsic local spin and charge fluctuations and facilitates the formation of local magnetic moments. 

To ensure accuracy, CMRT includes a sum-rule correction term that effectively redistributes non-local Coulomb interactions back into local terms but otherwise does not change the overall lattice Hamiltonian. The sum rule correction coefficients are provisionally represented as 

\begin{align}
	\lambda _{\mathit{ijkl};\sigma \sigma ^{'}}^{\mathit{\alpha \beta \gamma \delta }}=\lambda _{\mathit{i\sigma }}^{\alpha
	}\delta _{\mathit{ik}}\delta _{\mathit{jl}}\left(1-\delta _{\mathit{ij}}\right)\delta _{\mathit{\alpha \gamma }}\delta
	_{\mathit{\beta \delta.}}.
	 \label{ecmrC}%
\end{align}
In order to control the artifact caused by the Fock term out of the approximated two-body interaction treatment, the central term  $\lambda _{\mathit{i\sigma }}^{\alpha }$ in Eq. (\ref{ecmrC}) are defined for a paramagnetic state in this work as 
\begin{align}
\lambda _i^{\alpha }=\frac{\sum _{\sigma ^{'}}\left[\sum _{j{\neq}i}\sum _{\beta }U_{\mathit{ijij}}^{\alpha \beta
		\mathit{\alpha \beta }}\left|\left\langle c_{\mathit{i\alpha },\sigma ^{'}}^{\dag} c_{\mathit{j\beta },\sigma
			^{'}}\right\rangle \right|^2\right]}{\sum _{\sigma ^{'}}\left[\sum _{j{\neq}i}\sum _{\beta }\left|\left\langle
	c_{\mathit{i\alpha },\sigma ^{'}}^{\dag}c_{\mathit{j\beta },\sigma ^{'}}\right\rangle
	\right|^2\right]}.
\end{align}

The correlation matrix renormalization theory  is implemented by being interfaced with the Hartree-Fock (HF)
module of the Vienna \textit{Ab initio} Simulation Package (VASP)\cite{Kresse1996}. It utilizes the quasi-atomic minimal basis set orbitals(QUAMBO) \cite{Qian2008} to define its working local basis set. It achieves a computational speed comparable to a minimal-basis HF calculation\cite{Liu2021prb,Liu2021JPCM}, offering a significant advantage over quantum Monte Carlo methods.

The current CMRT calculation is a spin-polarized calculation without spin-orbit interaction effect included. The electronic structure calculations employed a kinetic-energy cutoff of 500 eV, and electronic self-consistency was ensured with a convergence threshold of $5\times10^{-5}$ eV. Brillouin-zone sampling was carried out using an automatically generated k-point grid corresponding to an  $R_K$ length of 30 in VASP. 
\section{Results and Discussion}

\subsection{Energetics and Phase Transition of CeN under Pressure}
\begin{figure*}
	\includegraphics[bb=50 30 680 520, width=8.0cm]{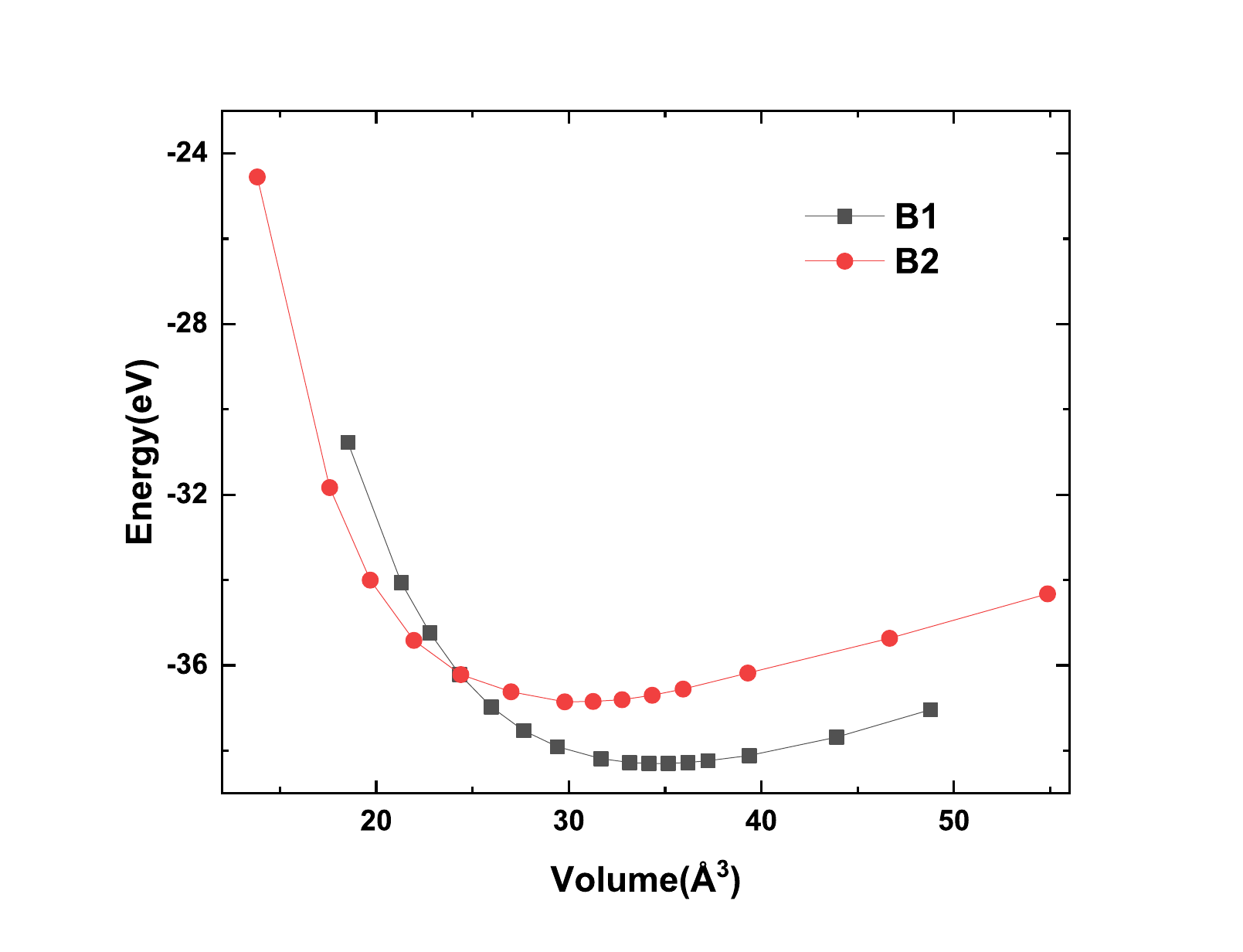}
	 \centering
	\caption{
		CMRT-calculated total energy of CeN as a function of unit-cell volume for the B1 (squares, black solid line) and B2 (circles, red solid line) phases.
	}
	\label{fig:ev-curve}
\end{figure*}

To evaluate the structure stability of CeN and predict its pressure-induced phase transition, we calculated the total energy as a function of volume for both the B1 (NaCl-type) and B2 (CsCl-type) phases, as shown in Fig. \ref{fig:ev-curve}. As expected, the B1 phase is energetically the more stable structure at ambient pressure. Our CMRT calculations reveal a well-defined energy minimum for the B1 phase at a lattice constant of 5.17  $\text{{\AA}}$, which is in good agreement with the experimental value of 5.02  $\text{{\AA}}$\cite{Olsen2012}. A Birch–Murnaghan Equation of State fit yields B $\approx$ 131 GPa for this phase, close to the measured values (156(3)\leavevmode\,GPa)\cite{Olsen2012}. The E-V curve for the competing B2 phase has its minimum located at a smaller volume but remains $\sim1.44$ eV higher in energy than B1 around their respective minima, establishing a large energetic separation between the two phases.
\begin{figure}
	\centering
	\includegraphics[bb=165 30 650 575, width=8.0cm]{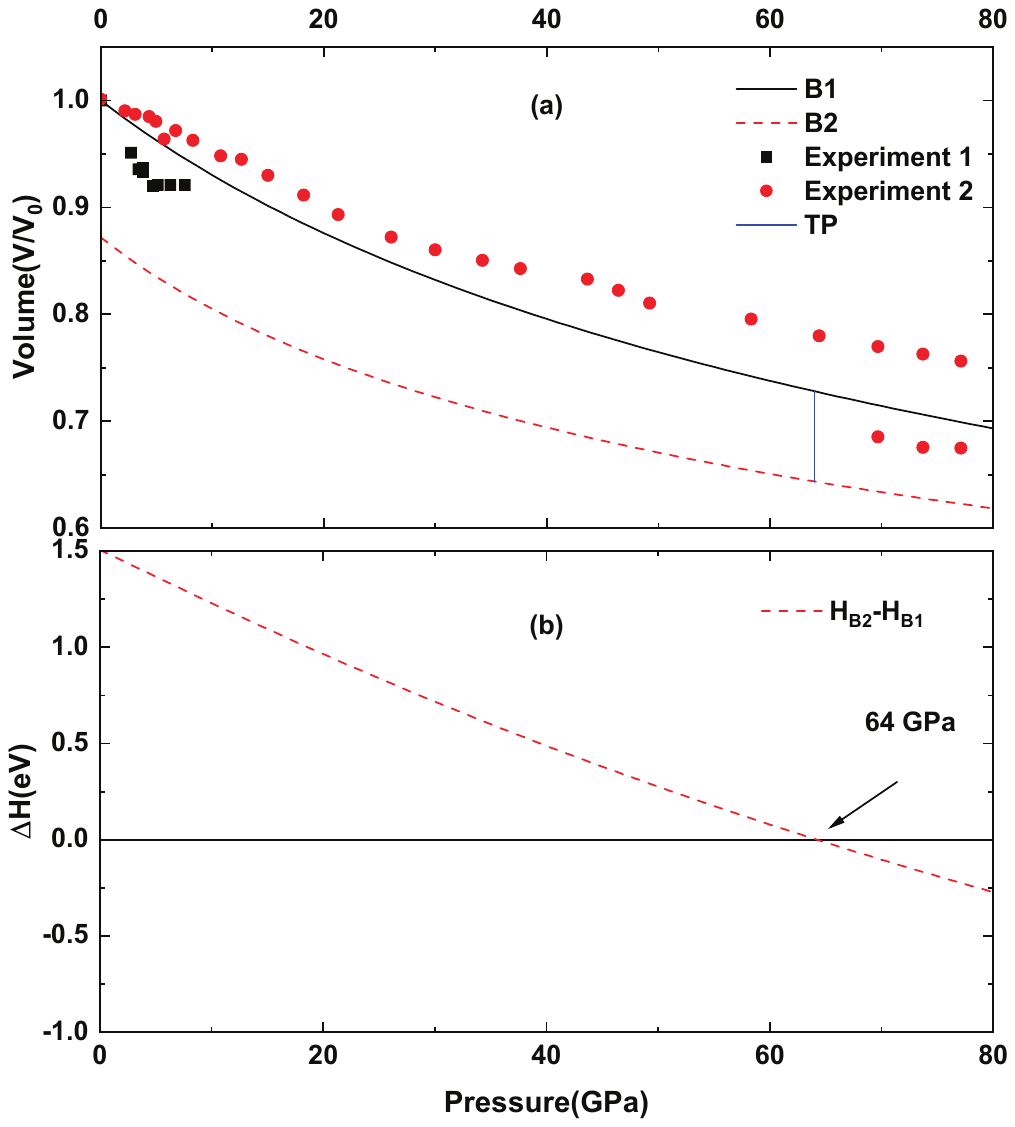}
	\caption{
		(a) Relative volume ($V/V_{0}$) vs Pressure (P) of CeN calculated using CMRT (B1 : Solid Line and B2: Dash line). $V_{0}$ denotes the equibrium volume of the B1 phase (at $P = 0$ GPa). The vertical blue solid line indicates the transition pressure (TP) corresponding to the B1 → B2 phase transformation. The points marked are the experimental data from Ref.12(black circle) and Ref. 23 (red circle); (b)Enthalpy differences and Estimation of phase transition pressure from B1 to B2 of CeN.
	}
	\label{fig:pv-Hp}
\end{figure}

To evaluate the structural phase transition pressure of CeN between the two phases, we calculated the pressure-volume (P-V) curves and evaluated the enthalpy difference for both the B1 (NaCl-type) and B2 (CsCl-type) phases. Our results are shown in Fig. \ref{fig:pv-Hp}. The calculated P-V relation (Fig. \ref{fig:pv-Hp}a) demonstrates a good agreement with experimental data \cite{Olcese1979, Olsen2012}, validating our computational approach. The curve shows a smooth compression of the B1 phase up to approximately 64 GPa, at which point an abrupt volume collapse of 11.5\% occurs as the structure transforms to
the denser B2 phase. This value closely matches the experimentally reported volume collapse of approximately
10.9(3)\% \cite{Olsen2012}. This structural transition is thermodynamically explained by the enthalpy difference, $\Delta H=H_{B2\hspace{0pt}}$-$H_{B1}\hspace{0pt}$, shown in Fig. \ref{fig:pv-Hp}b. At low pressures, the positive enthalpy difference confirms that the B1 phase has lower enthalpy and is thus thermodynamically more stable.
However, at 64 GPa, the enthalpy difference crosses zero, indicating that the B2 phase becomes energetically more favorable. This crossover correctly identifies the pressure at which the first-order phase transition occurs. Our calculated transition pressure of 64 GPa is in excellent agreement with the experimental range of 65–70 GPa \cite{Olsen2012}. These results demonstrate that CMRT accurately captures the intricate interplay between lattice mechanics and electronic correlations, providing a fully \textit{ab initio} and reliable description of the pressure-induced phase transition in CeN.

\subsection{Density of States Analysis}

\begin{figure*}
	\centering
	\includegraphics*[bb=55 10 800 550, width=8.0cm]{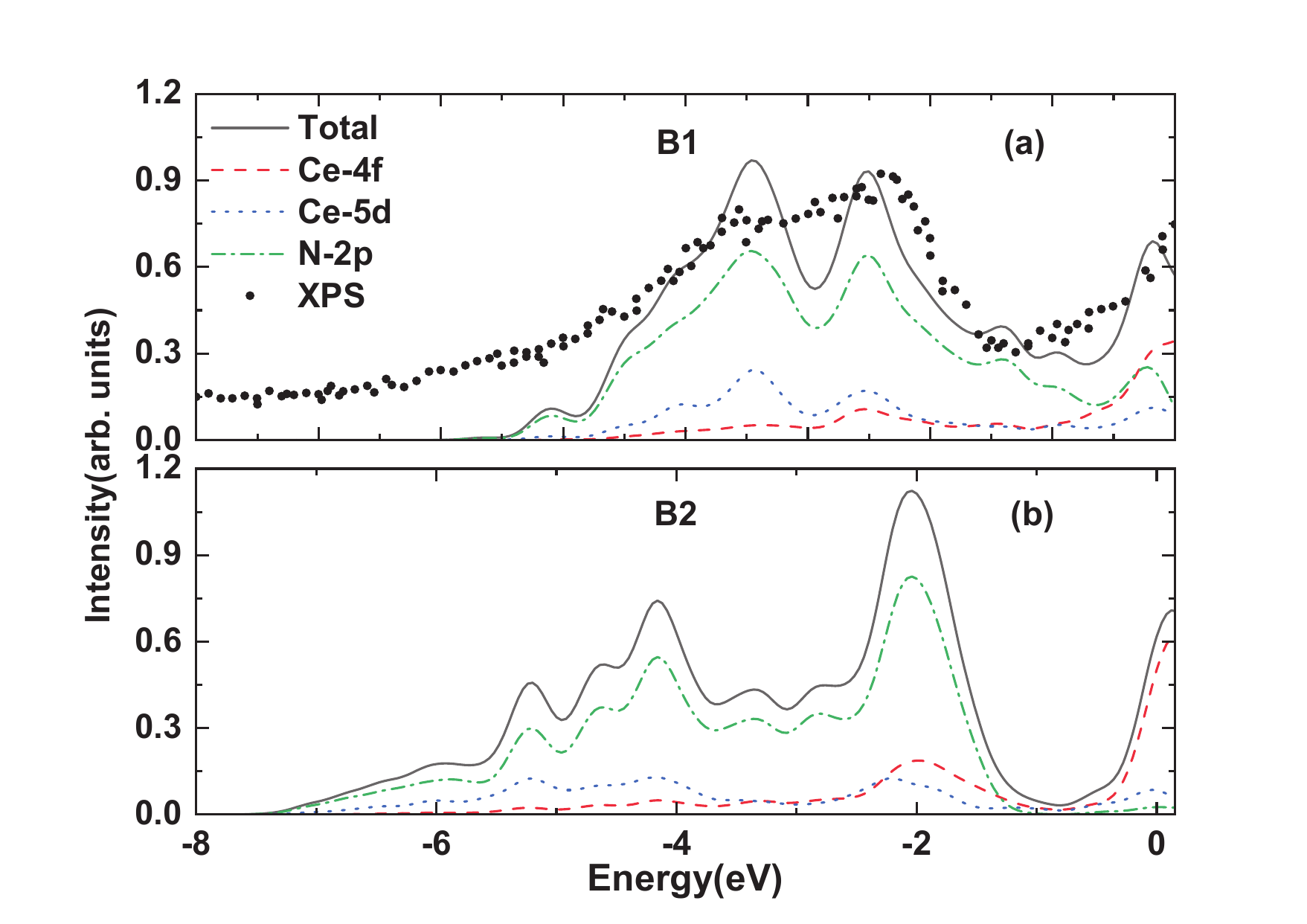}
	\caption{ Calculated spectra of B1-type(a) and B2-type(b) CeN. The experimental XPS spectrum reported in Ref. \cite{Baer1978} for the B1 phase is included in the upper panel for comparison, while the lower panel presents only the calculated results for the B2 phase. The calculated results include projections onto the Ce $4f$ and $5d$ states, as well as the N $2p$ state.
	}
	\label{fig:DOS}
\end{figure*}

\begin{figure}
	\centering
	\includegraphics[bb=80 40 740 565,width=8.0cm]{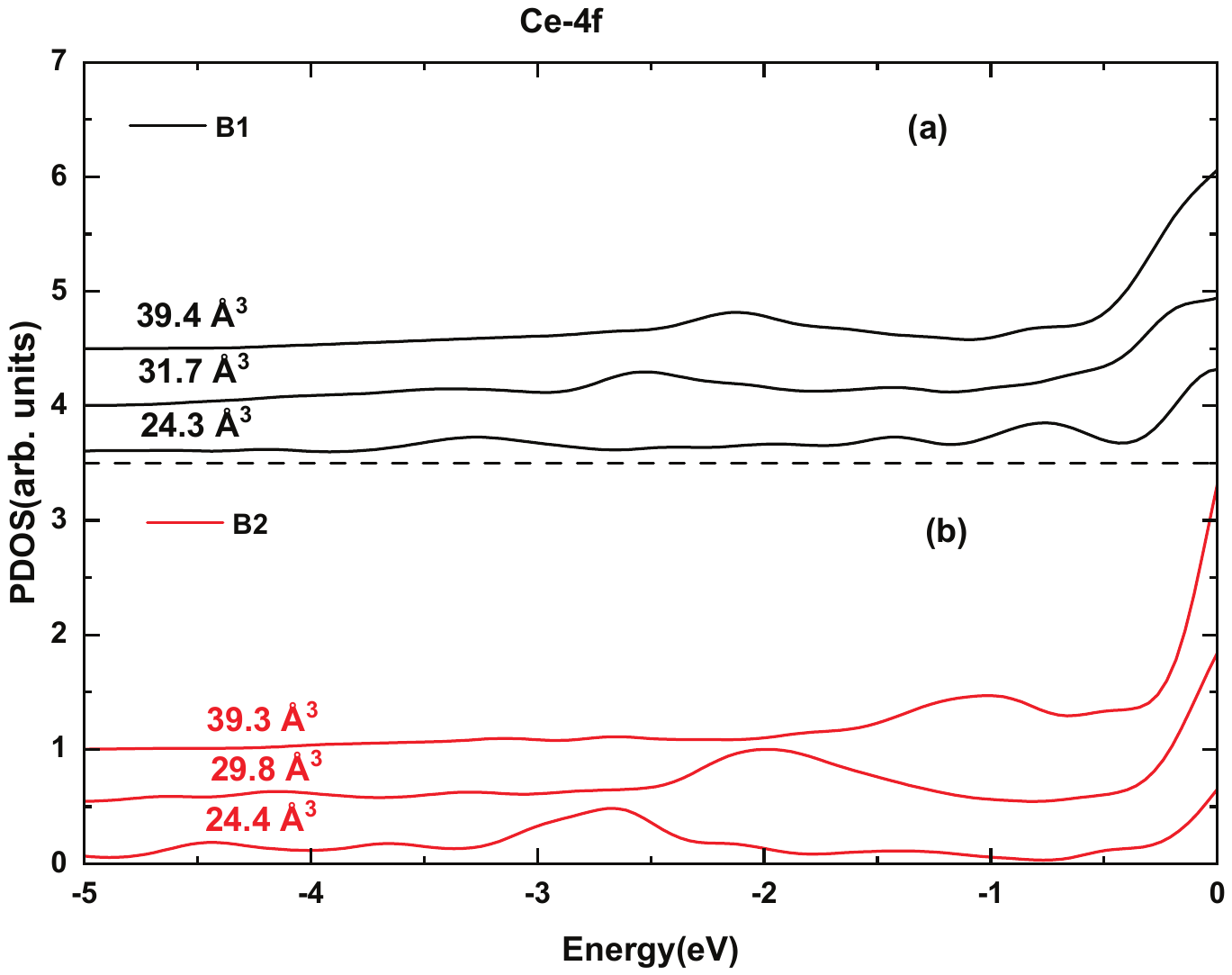}
	\caption{  Electronic density of $4f$ state at different volumes of CeN with the B1-type (a) and
		B2-type (b)  structures. }
	\label{fig:PDOS}
\end{figure}

To further understand the electronic mechanism underlying the pressure-induced B1 → B2 phase transition in CeN, we analyzed the total density of states (DOS, Fig. \ref{fig:DOS} and projected density of states (PDOS, Fig. \ref{fig:PDOS}) of $4f$-orbitals for
both structures. 
Figure \ref{fig:DOS} presents our CMRT-calculated valence-band spectra for the ambient-pressure B1 and high-pressure B2 phases of CeN. For validation, the experimental XPS spectrum reported for the B1 phase \cite{Baer1978} is also shown in the upper panel, whereas no experimental data are available for the B2 phase. The calculated spectra are projected onto Ce-$4f$, Ce-$5d$, and N-$2p$ orbitals and broadened to reproduce the experimental resolution. The energy scale is referenced to the Fermi level $E_F$\hspace{0pt}. For the B1 phase, the experimental spectrum shows a sharp, narrow peak pinned at  $E_F$\hspace{0pt}, accompanied by a broader incoherent hump, presumably the lower Hubbard subband, extending to higher binding energies. Our calculations reproduce these features and assign the low-energy resonance at Fermi level primarily to Ce-$4f$ weight with a smaller Ce-$5d$ admixture, while the deeper manifold at 3–5 eV below the Fermi level is dominated by N-$2p$ states with noticeable Ce-$5d$ hybridization. This orbital decomposition reflects the mixed-valence nature of CeN: the narrow $4f$
level remains partially occupied and pinned at  $E_F$, while hybridization with itinerant N-$2p$ and Ce-$5d$ states produces the broader valence band. Within the region (–3 to –5 eV), both experiment and calculation reveal a noticeable splitting feature around –3 to –4 eV. This splitting can be attributed to hybridization between the localized $4f$ state and the extended Ce–$5d$ and N–$2p$ bands. As pressure enhances these couplings in this mixed-valence system, a fraction of the $4f$ electrons become partially delocalized into the conduction manifold, strengthening the attractive potential in the valence region. As a result, the coupling between localized 4f states and extended Ce-$5d$ and N-$2p$ states modifies the band structure and gives rise to the observed splitting \cite{Baer1978}. This hybridization picture highlights the dual character of Ce $4f$ electrons in
CeN---partially localized yet strongly interacting with conduction bands---which is central to its intermediate-valence behavior and metallicity.

In the compressed B2 phase, both the total DOS and projected DOS exhibit an overall broadening due to enhanced hybridization between Ce-$4f$, Ce-$5d$, and N-$2p$ states. This broadening reflects the increased itinerancy of $4f$ electrons under pressure. Meanwhile, the $4f$-derived peak near the Fermi level remains sharp, which can be attributed to the higher $4f$ occupancy and the possible formation of a coherent quasiparticle resonance at  $E_F$. In other words, pressure drives a redistribution of spectral weight---spreading $4f$ contributions over a wider energy range while simultaneously reinforcing the coherent component at the Fermi level. Such coexistence of a broadened background and a sharp quasiparticle resonance signifies a correlated itinerant regime characteristic of the B2 phase.

Figure \ref{fig:PDOS} presents the Ce-$4f$ projected density of state at various volumes for both the B1 and B2 phases of CeN. It clearly shows that pressure strongly influences the electronic structure in both B1 and B2 phases, and this evolution plays a critical role in the pressure-induced phase transition of CeN. At the large volume (39.3 Å³), the B1 phase exhibits a pronounced $4f$ peak just below the Fermi level, indicating a more localized $4f$ character, while the B2 phase shows slightly broader features, suggesting a modest increase in itinerancy. As the volume decreases to 31.7 Å³ and further to 24.3 Å³, the B1-phase $4f$ peak near the Fermi level is progressively suppressed, whereas the B2-phase $4f$ states become more delocalized, with broader PDOS distributions and reduced peak intensity. This indicates enhanced $4f–spd$ hybridization in B2 under compression. The reduction in the sharpness of the $4f$ peak near the Fermi level with pressure, particularly in the B2 phase, signals a partial delocalization of $4f$ electrons, which may lower the total energy of the B2 structure relative to B1. This change in $4f$ electron behavior is an energetic driving force for the B1 → B2 transition, as the gain in kinetic energy from itinerant $4f$ states offsets the cost of structural rearrangement under high pressure.

\subsection{Mixed-Valence Behavior and Valence State of Ce in CeN}

\begin{figure}
\includegraphics[bb=40 200 500 580,  width=8.0cm]{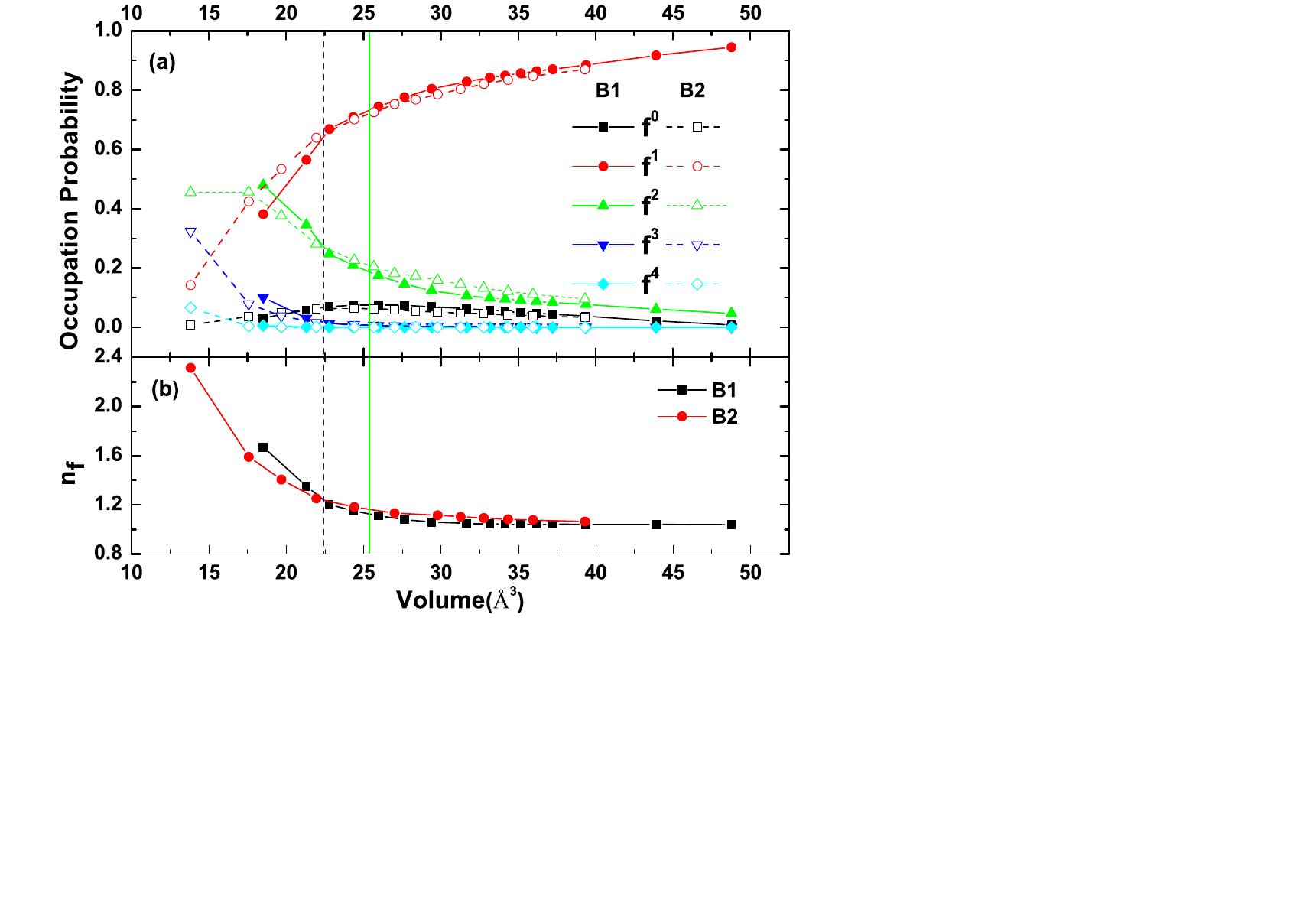}
\caption{(a) Occupation probabilities of $4f^{n}$ (n=0,1,2,3,4) configurations and (b) the
	number of occupied $4f$ electrons $n_f$  of B1-type and B2-type CeN as a function of the specific cell volume.
	The green solid and blue dash vertical lines indicate the cell volumes immediately before and after the B1→B2 transition,
	respectively.
 }
\label{fig:occupation-nf}
\end{figure}

The valence state of cerium in CeN has long attracted interest due to the dual nature of its $4f$ electrons. In this study, CMRT calculations were performed to determine the evolution of the $4f$ electronic configurations and the corresponding $4f$ occupation number ($n_{f}$) as a function of the specific cell volume for both B1- and B2-type CeN, as illustrated in Fig. \ref{fig:occupation-nf}. The calculations results indicate a dominant $4f^{1}$ occupation of approximately 0.857 for B1-type CeN at its equilibrium volume, in close agreement with the experimental value of $\sim0.83$ obtained from the Gunnarsson-Schönhammer(GS) fitting of XPS/BIS measurements \cite{Wuilloud1985}. It should be noted that in their model calculation, only the $4f^{0}$ and $4f^{1}$ configurations are
considered, which results in $n_{f}=p_{f1}$ \cite{Wuilloud1985}. This strong agreement supports the reliability of the calculated $4f$ occupation and reinforces the picture of CeN as a mixed-valence system, with Ce exhibiting a predominantly trivalent ($4f^{1}$) character.

The total $4f$ occupation number of B1-type CeN, derived from the CMRT probabilities ($p_{f^0}=0.051$, $p_{f^1}=0.857$, $p_{f^2}=0.091$) is $n_f=p_{f^1}+2p_{f^2}=1.04$ \cite{Gunnarsson1983}. This corresponds to a formal Ce valence of approximately  $4-n_f=+2.96$, slightly below the ideal trivalent value. This deviation from a pure Ce$^{3+}$ configuration arises naturally from the finite $f^{2}$ contribution. Each $f^{2}$ component introduces one additional $4f$ electron beyond the $f^{1}$ configuration, thereby increasing the average $4f$ occupancy and reducing the effective Ce valence. Although the $f^{1}$ configuration dominates ($\sim85.7\%$), minor admixtures of $f^{0}$ and $f^{2}$ configurations slightly reduce $n_f$ from unity, indicating a predominantly trivalent Ce with weak $4f$–conduction hybridization, as supported by the density-of-states analysis. The calculated effective valence ( $+2.96$) is in good agreement with spectroscopic estimates (${\approx}$+3.1 to +3.2) obtained from XPS/BIS\cite{Wuilloud1985} and DMFT analyses \cite{Liu2024,Li2023}, which likewise identify CeN as an intermediate-valence compound. In this regime, Ce$^{3+}$ dominates while a small Ce$^{4+}$ component emerges, reflecting the delicate balance between $4f$ localization and itinerancy. These valence fluctuations likely drive CeN$'$s unusual magnetic and transport behaviors under varying pressure and temperature.

Analysis of different $4f$ configurations reveals that, under compression, the proportion of the $4f^{1}$ configuration
decreases in both B1 and B2 structures, while $4f^{0}$ and $4f^{2}$ configuration contributions increase(Fig. \ref{fig:occupation-nf})(a)). This trend
reflects progressive delocalization and thus mobility of Ce $4f$ electrons, enhanced hybridization, and the emergence of
valence fluctuations, marking a gradual transition trend towards a more itinerant $4f$-electron behavior. In the B1 phase, the decrease of the $4f^{1}$ configuration fraction accelerates markedly as the volume approaches the transition region, and this rapid suppression is more pronounced than in the B2 phase. Correspondingly, the probabilities of the other configurations (mainly $4f^{0}+4f^{2}$) rise more steeply in B1, signaling a stronger pressure-driven enhancement of charge fluctuations and $4f$–spd hybridization during the final stage of compression. In contrast, the B2 phase exhibits a more gradual evolution of all configuration weights, consistent with its higher coordination environment and its already more itinerant $4f$ character. The sharp contrast between the steep redistribution in B1 and the smoother variation in B2 highlights the distinct electronic responses to compression in the two structures. This evolution of the $4f$ configuration weights provides an electronic-structure perspective that aligns with the anomalies observed in our calculated $E$–$V$ relation. Although finite-temperature vibrational contributions are not included—and thus the present calculations do not yield true free energies—the pronounced change in the slope of the zero-temperature $E$–$V$ curve near the transition volume effectively serves as the enthalpic signature of the B1 → B2 structural transformation.

We further examined the evolution of the total $4f$ occupation number  $n_f$ \ for both B1 and B2 phases as a function of
cell volume (Fig. \ref{fig:occupation-nf}(b)). As the volume decreases under compression,  $n_f$ \ increases steadily from near unity,
indicating growing participation of $4f$ electrons in bonding. Interestingly, the  $n_f$ curves for B1 and B2 intersect
within the transition region.  For volumes larger than those in the transition region, the B2 phase displays a higher $n_f$ than B1, whereas for volumes smaller than this region, B2 shows a lower $n_f$.  This crossover highlights the electronic reconstruction accompanying the
structural transition.

Collectively, these results demonstrate that the structural transition in CeN is strongly coupled to the $4f$-electron
reorganization. In the large-volume regime, B1 behaves closer to trivalent Ce, while B2 shows enhanced electron
transfer into the conduction band. Under high pressure (small volume), the $4f$ electrons in B2 become increasingly
delocalized and hybridized, actively contributing to the electronic band reconstruction. Thus, the B1 → B2 transition
is not purely structural but is intimately linked to $f$-electron delocalization and valence fluctuations. This mechanism
is reminiscent of heavy-fermion systems, where pressure-induced valence changes drive abrupt structural
transformations \cite{Temmerman2006}.

\section{Conclusion}
In summary, we have investigated the pressure-induced structural and electronic evolution of CeN using correlation
matrix renormalization theory in direct comparison with experimental spectroscopies. Our results show that the
ambient-pressure B1 phase exhibits a mixed-valence 4f character, reflecting strong screening effects and significant
$p–d–f$ hybridization between the 4f and conduction states. Under compression, the transition to the B2 phase is
accompanied by a redistribution of spectral weight: the 4f band broadens, hybridization with Ce-$5d$ and N-$2p$ states is
enhanced, and the effective $4f$ orbital occupancy increases. These features signify a crossover toward a more itinerant
$4f$ character that underpins the ${\sim}$11\% volume collapse at the B1 → B2 transition. Moreover, the present study
clarifies the dual role of Ce $4f$ electrons---partially localized yet strongly
hybridized---in driving the correlation-induced structural instability of CeN. 

Beyond qualitatively reproducing experimental XPS spectra and capturing the mixed-valence fingerprints of CeN, CMRT
provides a consistent description of both total energies and spectral properties without adjustable parameters. While a
comprehensive study of this system calls for inclusion of spin-orbit coupling interaction and possibly improved
numerical treatment within CMRT to fully study the fine $4f$ multiplet structures and to be closer to the experimental $4f$
electron occupation. While details of the results might dependent on spin-orbit coupling interaction added to the CMRT,
the fact that the current CMRT study has already been able to capture most eminent experimental features of this
material establishes CMRT as a valuable many-body approach for $f$-electron systems, complementary to DFT+DMFT but
computationally more efficient.

\section*{Acknowledgement}
Work at Ames National Laboratory was supported by the U.S. Department of Energy (DOE), Office of Science, Basic Energy
Sciences, Materials Science and Engineering Division including a grant of computer time at the National Energy Research
Scientific Computing Center (NERSC) in Berkeley. Ames Laboratory is operated for the U.S. DOE by Iowa State University
under Contract No. DE-AC02-07CH11358.

\bibliography{CeN}
\end{document}